ARTICLE

# Do Cell Phones Cause Cancer?

BY BERNARD LEIKIND

News reports threaten that our cell phones may cause cancer—brain cancer, eye cancer, and others. We are told that fragile children's developing brains are at risk. Concerned epidemiologists collect their data and warn that they cannot rule out the possibility of harm from cell phone radiation and that they must do more research. Medical professionals assert, as a precaution and in the absence of definitive data, that we should place our phones at arm's length. News accounts fill us with alarm. Danger lurks.

Fears that cell phones cause cancer are groundless. There is not a shred of evidence that the electromagnetic radiation from your cell phones causes harm, much less that from the wiring in the walls of your house, your hair dryer, electric blanket, or the power distribution wires nearby.

We know exactly what happens to energy from any of these sources when it meets the atoms and molecules in your body, and that energy cannot cause cancer. There is no known way that this energy can cause any cancer, nor is there any unknown way that this energy can cause any cancer.

There is a link between some forms of electromagnetic radiation and some cancers. These forms of electromagnetic radiation are ultraviolet radiation, X-rays, and gamma rays. They are dangerous because they may break covalent chemical bonds in your body. Breakage of certain covalent bonds in key molecules leads to an increased cancer risk. For example, there is a link between ultraviolet light from the sun and skin cancers.

All other forms of electromagnetic radiation other than these may add to molecules' or atoms' thermal agitation, but can do nothing else. Visible light has sufficient energy to affect chemical bonds. When light strikes the cones and rods in our retinas rhodopsin bends from its resting state to another, but it does not break. When visible light strikes the chlorophyll molecules in plants, electrons shift about but the chlorophyll does not break. Visible light does not cause cancer.

Electromagnetic radiation transfers its energy to atoms and molecules in chunks called *photons*. The energy of a single photon is proportional to the photon's frequency. The photons of high frequency radiation, such as ultraviolet light, X-rays, and gamma rays, carry relatively large amounts of energy compared to those of lower frequency radiation. That is why high-energy photons can break covalent chemical bonds while the photon energy of all other forms of electromagnetic radiation, including visible light, infrared light, microwave, TV and radio waves, and AC power cannot.

*Figure 1* shows a range of energy that is important for life and for the science of biochemistry. The figure displays an energy scale to help you place relevant energy states or processes in context. Horizontal positions indicate the energy range.

Look at the area covered by the long bracket in the middle. It shows the general energy range of the major strong chemical bonds—covalent bonds—which are significant for all of life's molecules. Below to the right you can see where the energy of an important organic covalent bond—that which occurs between two carbon atoms—falls on the scale. Further up the scale, on the upper right, is the energy range of carcinogenic electromagnetic radiation. Notice where the call out for green light falls on this scale. Visible light does not cause cancer.

Notice that the energy of cell phone radiation and AC power radiation in this scale is very low. Cell phone radiation cannot break, damage, or weaken any covalent bond.

*Figure 2* shows the lowest energy part of *Figure 1's* energy scale. *Figure 1* ranges from 0 kJ/mole to 600 kJ/mole. *Figure 2* ranges from 0 kJ/mole to 30 kJ/mole.

Notice the bracket that shows the range of weak bonds in each figure. These are hydrogen bonds, van der Waals bonds, electrostatic bonds, and various other effects, such as hydrophobic or hydrophilic forces. In the complex molecules of life, these bonds play critical roles holding strands together and creating the three-dimensional shapes of molecules.

Covalent bonds hold together the single strands of DNA. Hydrogen bonds connect one strand to its mate. Enzymes fold and twist to create the forms they require as they perform their role as catalysts. The various weak bonds maintain the shapes of these folds and twists.

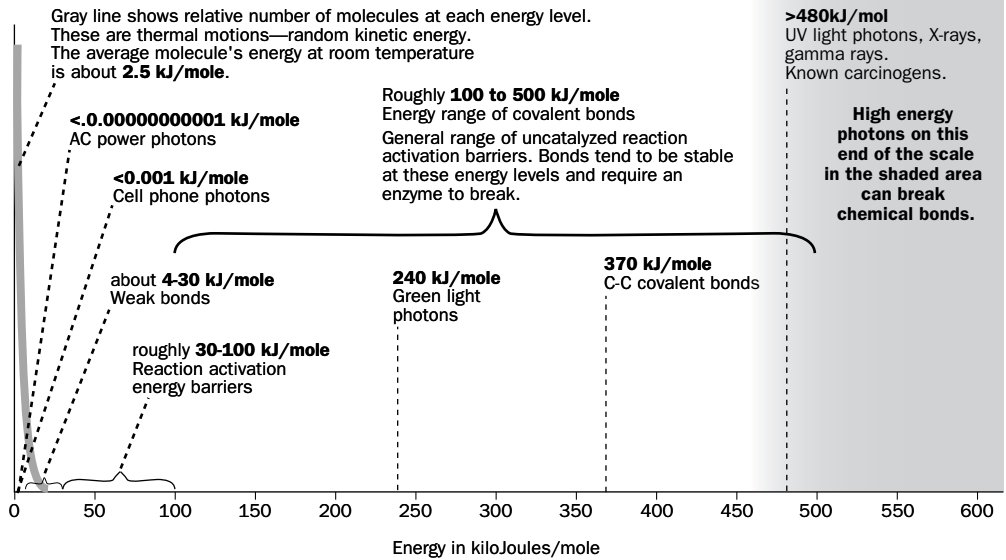

Figure 1. The units of this scale are familiar to chemists. Chemists like to think about test-tube-sized quantities of stuff. A mole is a unit that measures how much stuff you have. It is a count of objects: atoms, molecules, photons, chemical bonds. One mole of any object contains $6.023 \times 10^{23}$ of those objects. Physicists prefer to state the energy in one bond or in one photon. A physicist would divide all the numbers in this figure by the number of objects in a mole to show the energy in Joules in a single object. An (old) physicist might prefer to express this energy in units of electron volts. Measured in electron volts, the energy in one green light photon is about 2.5 electron volts. The energy in one banana is 150 to 200 Calories, which corresponds to 600 or 800 kJ/banana; that is, one banana, not a mole of bananas.

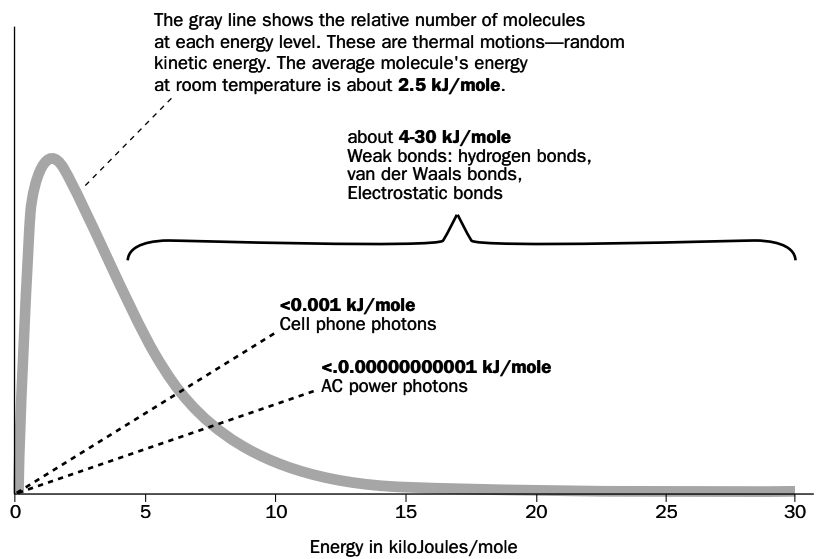

Figure 2.

Drawn in both figures is a graph that suggests the energy of molecular thermal motions at body temperature. Everything in our bodies partakes in these thermal motions. The molecules jostle one another. They twist and vibrate. The thick grey line on the graph shows how energy distributes itself among these various motions. The motion's average energy is about 2.5 kJ/mole. Some molecules, but not many, have much more energy.

If energy transfers of 2.5 kJ/mole, more or less, were sufficient to damage life's molecules, life would be impossible because random thermal motions would quickly break most of them. Fortunately, covalent bonds require ten to fifty times this amount of energy transfer before they break. Thermal jostling does not interfere with them. Weak biochemical bonds, however, live within the upper range of thermal bonds and shakes. That is why they do not enter into life's structure as single bonds, but always as groups. In the long double helices of DNA, the hydrogen bonds are like the individual teeth of a long zipper. Together they withstand what any single one of them could not.

These collisions are electromagnetic interactions. The molecules' outer electrons sense the presence of their neighbors though electromagnetic forces. These electrons resist oncoming neighbors, pushing them away, and pushing upon their own molecules as well. Electromagnetic forces transmit these pushes. All of the molecules of biology must be able to withstand these electromagnetic forces to maintain their shapes and their functions. The forces that electromagnetic fields from cell phones exert on life's molecules are no different from any of these molecular pushes, except that they are much, much smaller.

Cancer is a disease of the heredity of individual cells. Something must cause a cell to begin transferring mistakes to its progeny. One cell goes haywire, replicating wildly, transmitting the mistaken instructions—the damaged DNA—to each of its daughters. If the damage is too great, the cell will die. If the damage is not sufficient, it is not cancer. The damaged cell and its damaged progeny must continue to function in their crippled, uncontrolled states. Cancer generally requires more than one mutation in a single cell.

It is worth understanding how chemical changes occur, why life's molecules are stable in the cytoplasm, and how life controls its chemical reactions, turning them on or off. Consider *Figure 3*.

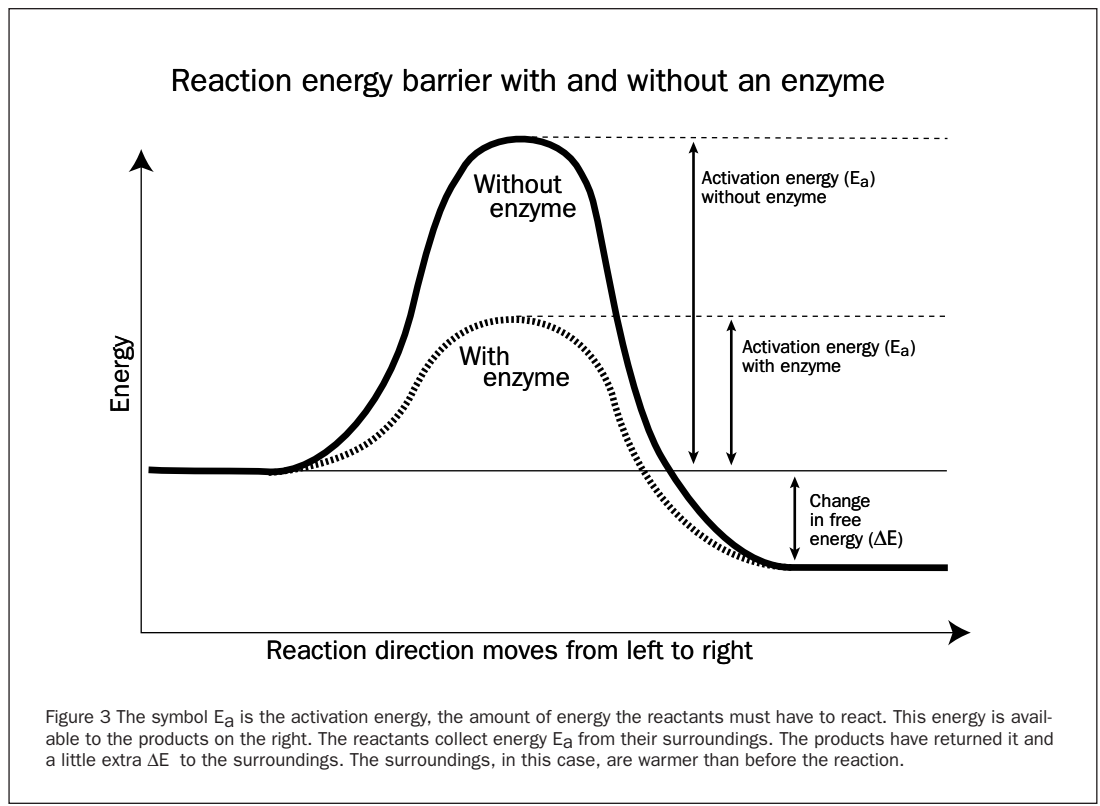

Figure 3 The symbol $E_a$ is the activation energy, the amount of energy the reactants must have to react. This energy is available to the products on the right. The reactants collect energy $E_a$ from their surroundings. The products have returned it and a little extra $\Delta E$ to the surroundings. The surroundings, in this case, are warmer than before the reaction.

This famous diagram appears in all biochemistry books. It is a schematic representation of a reaction. Consider this reaction A + B → C + D, where A and B are reactants and C and D are products. In the diagram and the equation, the reaction begins on the left and moves to the right. The vertical scale is energy. Don't worry about the technical details. Begin with the upper solid black line with the label *Reaction Energy Barrier without an Enzyme*. For this reaction, the molecules A and B must assemble sufficient energy to carry them over the hill. This energy may come from the incessant thermal collisions, from some other molecule's internal energy, from an incoming photon of electromagnetic radiation, or other sources. The total energy of the entire system, including the surroundings, is a constant.

Through the continual random exchange of energy between the molecules A and B and their surroundings, if A and B happen to meet when they have sufficient energy to make it over the top of the hill, then they will react, forming C and D. These products appear on the diagram's right.

This diagram is illustrative. The actual diagram of even a simple reaction might have several dimensions in place of the single horizontal axis. The hills would be complicated surfaces with mountains and valleys. The diagram would have to take into account factors such as the orientation of the reactant molecules, and much else. It is the case, however, that all of life's stable molecules live in a well—a valley—similar to the left side of the diagram. They will require an injection of energy from their surroundings to escape. Biological molecules have many possible reactions in which they might take part. Remove an atom and replace it with another. Switch any molecular piece with another molecular piece. Natural selection has designed all of the molecules of life so that they are stable in chemical composition, form, and function. High activation energy barriers in all directions make all possible reactions rare. If this were not the case, then the molecules of life would not be stable.

When life requires a particular reaction to take place, there will be an enzyme to facilitate it. An enzyme is a biological catalyst. Consider the lower dashed line in *Figure 3*. This line has the label *Reaction Energy Barrier with an Enzyme*. This depicts the same reaction A + B → C + D, but this time there is an enzyme to facilitate the reaction. Without going into the remarkable details of enzymatic function, we can say that the enzyme has the

## A Cell Phone's Heating Power

The central premise of this paper is that the only effect cell phone radiation *can* have on our bodies is to warm them. Consider this. I am about the same height (6' 1") and weight (185 pounds) as President Obama. My basal metabolic rate is about 1750 Calories per day. This is the energy my body uses just to keep me idling at the desk. Add in 250 Calories each day to account for mowing the lawn and vacuuming the rug. The average energy I might generate each day is about 2000 Calories.

Calories per day, energy per unit time, represents power. Convert 2000 Calories per day into physics units, Watts or Joules per second. The result is about 100 Watts. Thus, as I go about my ordinary life I am the power equivalent of a 100 Watt light bulb on all the time.

Many days, I visit a health club and jog on one of their treadmills. I like the level 2 hill workout, and I clomp along at about a 9-minute mile pace for 30 minutes. According to the treadmill's display, this burns about 500 Calories. Five hundred calories in 30 minutes corresponds to a power of 1150 Watts. This is 11 or 12 times my usual power; 11 or 12 100-Watt light bulbs on for 30 minutes in my leg muscles.

The efficiency of the human body as it converts internal power into external work is a complicated matter that depends upon many factors. For the purposes of this estimate, it would not be misleading to take that efficiency to be about 20%. Thus, of the 1150 Watts I am using as I jog, about 230 Watts go to keeping me on the treadmill, and about 920 Watts go into heat in my leg muscles. Blood flows through those muscles bringing in oxygen and fuel and carrying away carbon dioxide and other waste products. The blood also warms to the temperature of the muscles and carries that energy throughout my body. My core body temperature rises, and I sweat a lot.

A cell phone radiates a Watt or two of electromagnetic radiation. Most of that goes out in every direction and some makes it to the cell phone tower. My body absorbs some of it in my hand, my ear, my skull, my brain tissues, and so on. Let's say that my body absorbs one Watt of this power in those nearby tissues. Those tissues warm a bit, and the blood flowing through them warms too. The blood carries any extra heat energy throughout the rest of my body, which it eventually transfers to the air around me.

No one believes that the 900 Watts that I generate in my leg muscles during hard exercise causes cancer in those muscles or anywhere else in my body. Why would anyone believe that 1 Watt of heating power from a cell phone might cause cancer?

effect of lowering the activation energy barrier for the reaction. With lower activation energy, the thermal jostling or other sources of reaction energy have a much easier time pushing the reactants over the hill. The reaction rate goes from nearly zero to some reasonable value.

There are no enzymes for unwanted reactions. Enzymes have and maintain the proper constitution and form to work correctly. For a mutation to occur or an enzyme to change, the energy for the chemical reaction must come from some place. An X-ray photon—from a cosmic ray, from the earth's radioactivity, or from an X-ray machine—may provide the required energy. Photons from any other form of electromagnetic radiation cannot.

Glance at *Figure 1* again. All of those chemical bonds across the middle of the diagram are stable. They do not break and reform, unless there is an enzyme to do it. On the left of the diagram is the graph showing the energy available from ordinary thermal motions to break these bonds. Also on the left is a bracket showing the range of typical activation energies. Thermal motions are insufficient to take molecules over the activation energy barrier for any reaction. Far to the right, however, you can see the photons of ultraviolet light, X-rays, and gamma rays. These photons may break bonds. They may cause mutations directly. They may damage individual enzyme molecules. Even green visible light photons in the middle of this range do not have enough energy to break bonds and take molecules over any activation energy barrier.

Now find the photons of cell phone radiation and of AC power. They are at the far left of this diagram. No photon from a cell phone can ever break a chemical bond. Making the radiation more intense does not make the photons stronger. It just means that there are more of them. The photons cannot gang up. Lots of them cannot do what one of them cannot.

When those weak photons disappear into a molecule, the molecule shifts and quivers a tiny bit. Its energy is a bit larger and the photon is gone. The molecule adjusts itself to its new slightly higher energy, and in subsequent collisions with its neighbors, it may transfer some of that energy to them. The temperature of the biological soup—the cytoplasm—is then a bit higher. The amount of heating due to cell phone radiation is small compared to your microwave, or standing in the sunshine, or wearing a scarf around your neck. This small increase in temperature does not cause cancer.

If cell phone photons or AC power photons, far to the left of *Figure 1*, were able to cause cancer by any mechanism, known or unknown, then those thermal vibrations also shown to the left side of the diagram would also cause cancer. So would all the forms of electromagnetic radiation that have more energetic photons than cell phone radiation.

Some of the concern over cellphone radiation may have originated from the normal statistical fluctuations that occur when studies are conducted. In recent years, epidemiologists have found significant environmental hazards, such as smoking and asbestos. They are now searching for hazards among much weaker effects. Some studies of a supposed hazard will show a small risk. Others studies of the same hazard will show no risk. In fact, some studies of the same potential hazard will show a benefit. This is the sign that there is no hazard, only statistical fluctuations. But only the studies that suggest risks, even small risks, will make news.

We can all be confident that any epidemiological study that purports to show that cell phone radiation causes any cancer must have at least one mistake. We can be certain because there is no plausible—or even implausible—mechanism by which cell phone radiation can cause any cancer.

When asked for a physicist's advice about cell phone safety, I explain that the radiation cannot cause cancer by any mechanism, known or unknown. If I am further pressed for comment I respond, "don't text while you drive, and don't eat your cell phone." 🆂

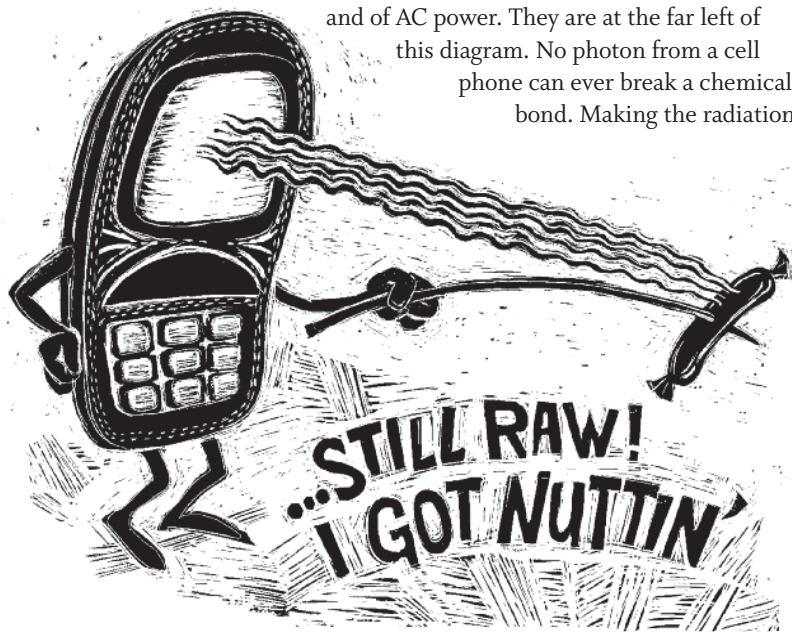

*Acknowledgment: I thank physicists Dr. Arthur West and Dr. Craig Bohren, and biochemists Dr. Joseph H. Guth and Dr. Jill Ferguson, for their careful review of this paper and for their suggestions.*

**Do Cell Phones Cause Cancer?**

Bernard Leikind

http://www.skeptic.com/eskeptic/10-06-09/

**Executive Summary:** Do cell phones, household electrical power wiring or appliances, or high voltage power lines cause cancer? Fuggedaboudit! No way! When pigs fly! When I'm the Pope! Don't text while you're driving, however, or eat your cell phone.

Microwave radiation from cell phones cannot cause cancer by any mechanism, known or unknown. My answer to the question in the title of this essay is no. Really, my answer is **NO!!!**

This essay is a companion to my article of the same title that appears in **The Skeptic,** Vol. **15**, no. 4. Here I will describe what all physicists know to be true about what happens when human tissue or any material absorbs microwave radiation. It is this knowledge that leads me to assert with such vehemence that cell phones do not cause cancer. I will also consider two recent, major epidemiological studies from Europe that correctly showed that there was no relationship between cell phones and brain cancers. It is a remarkable fact that the researchers, epidemiologists, evidently expected that their results would link cell phone use and brain cancer.

The considerations of this essay and my Skeptic article apply equally to 60 Hz, AC power from the wiring in our walls, from our hair driers, electric blankets, or televisions, and from high voltage power lines. None of these causes cancer either.

A cell phone emits about 1 Watt of electromagnetic radiation. Most of that zooms away to find a cell phone tower. The tissues of the user will absorb a part of this radiation. These tissues include the caller's hand, ear, scalp, skull, and brain. The closer a tissue is to the cell phone's antenna, the more of the radiation the tissue absorbs. For some reason, however, none of those raising fears about cell phones causing cancer are concerned about skin cancers on palms, fingers, or ears.

The frequency of the typical cell phone radiation is about 2.5 GHz, two and a half billion flips back and forth per second. The radiation travels at the speed of light, 186,000 miles per second, and dividing the one by the other and correcting for the units I used for the speed, shows that the wavelength of this radiation is about 10 centimeters or about 4 inches.

As the electric fields of the waves pass through the body's tissues, the fields grab and try to shake any molecules or parts of the molecules that they can. These fields like to grab and shake water molecules, and there are plenty available. The fields will grab whatever else they can, which may be all or of parts of many of the critical molecules of biochemistry, such as the DNA in genes, or enzymes, fuel molecules, waste molecules, structural molecules, and so on.

All of these molecules exist within the cytoplasm, and they are in close touch with one another. The molecules are quivering, twisting, and shaking, rattling about and transferring energy between each other. During the time, less than one billionth of a second, that it would take the cell phone's radiation to shake a molecule or part of a molecule back and forth, that molecule will suffer a thousand or ten thousand collisions with its neighbors. Any energy that the one molecule might begin to gather from the electromagnetic field rapidly spreads throughout all of its neighbors.

Coursing nearby to these molecules is a capillary filled with blood plasma and blood cells. This blood is at body temperature. Any extra energy from any source that appears in cells close to the capillaries will transfer into the slightly cooler blood, warming it. The flowing blood will carry the energy throughout the body. The body temperature will increase imperceptibly, and the extra energy will eventually transfer from the skin into the environment.

I have been expressing these ideas freely and non-technically. The ideas, however, are accurate, and I could have expressed them with long, technical or obscure words, as medical or scientific professionals will acknowledge. Physicists know precisely and in detail exactly what happens to every bit of the energy that leaves the cell phone and is absorbed by the body's cytoplasm. Anyone who asserts that cell phone radiation causes cancers must begin with this process.

Anyone who puts forward a potential mechanism by which this energy flow, less than 1 Watt, might cause any cancer should notice that he has thereby explained too much. One watt is much smaller than many other natural energy flows that no one suspects might cause cancer. In my Skeptic paper, I show that the average energy production in my body as I go about my life is about 100 Watts. I also show that while I jog on my local gym's treadmill for half an hour, I produce 1100 or 1200 Watts. This energy, produced in my leg muscles, travels throughout my body including my brain, and I sweat a lot. My body's temperature does not change much. No one believes that my frequent treadmill sessions cause cancer. If the cell phone's less than 1 Watt causes cancers, then why doesn't my exercise session's more than 1000 Watts cause cancer?

Within the past year the results from two major epidemiological studies appeared in the scientific literature and to great fanfare in the media. Plainly stated, these two different kinds of studies found no evidence to link cell phones and brain cancers. The researchers might have simply said, "We did these large, carefully designed studies, and cell phones have nothing to do with brain cancer."

In the major Danish study, the researchers collected data from the entire populations of Denmark, Sweden, Norway, and Finland. These sensible countries have long provided medical care for all of their fortunate residents. Therefore, the researchers had access to thorough records. Brain cancers are rare, so they must search through large populations to find sufficient cases to draw conclusions. The plan of this study was to compare trends in the incidence of brain cancers from the late 1980s into the mid 1990s when cell phone use was non-existent or rare with the incidence in the first decade of the 21$^{st}$ century when cell phone use was wide spread. They saw no effect. None. Zero. Nada. Zilch. Negative.

How did these researchers explain this result? These researchers believe that cell phones must cause brain cancer somehow to some degree. Therefore, they asserted that perhaps their study was not large enough, perhaps their study did not cover sufficient time, or perhaps the large sample population diluted the effect in susceptible subgroups. They grudgingly admitted that it was possible that their study showed no effect because cell phones do not cause cancer.

The other study, known as the Interphone study, is a case-control study. Searching the populations of 13 European nations the researchers found 6000 brain cancer patients. Next, the researchers sought out 6000 more people to form a matched control group. Then the epidemiologists searched their data to see if they could detect suggestions that cell phone use might increase the risk of brain cancer.

Here are a few quotations from the researchers about their results from a Reuters report:

> "The results really don't allow us to conclude that there is any risk associated with mobile phone use, but... it is also premature to say that there is no risk associated with it," the IARC's director Christopher Wild told Reuters.

Also:

> Data from the IARC study showed that overall, mobile telephone users in fact had a lower risk of brain cancer than people who had never used one, but the 21 scientists … said this finding suggested problems with the method, or inaccurate information from those who took part.
>
> Other results showed high cumulative call time may slightly raise the risk, but again the finding was not reliable.

> "We can't just conclude that there is no effect," said Elisabeth Cardis of the Centre for Research in Environmental Epidemiology in Barcelona, Spain, who led the study.
>
> "There are indications of a possible increase. We're not sure that it is correct. It could be due to bias, but the indications are sufficiently strong... to be concerned."

Why aren't these researchers proclaiming the brilliant discovery that cell phones protect against brain cancer? Why do they believe that concern is justified? They are confident that there is no possible way for cell phones to reduce the risk of brain cancer, but they suspect that the physicists might be wrong that there is no mechanism.

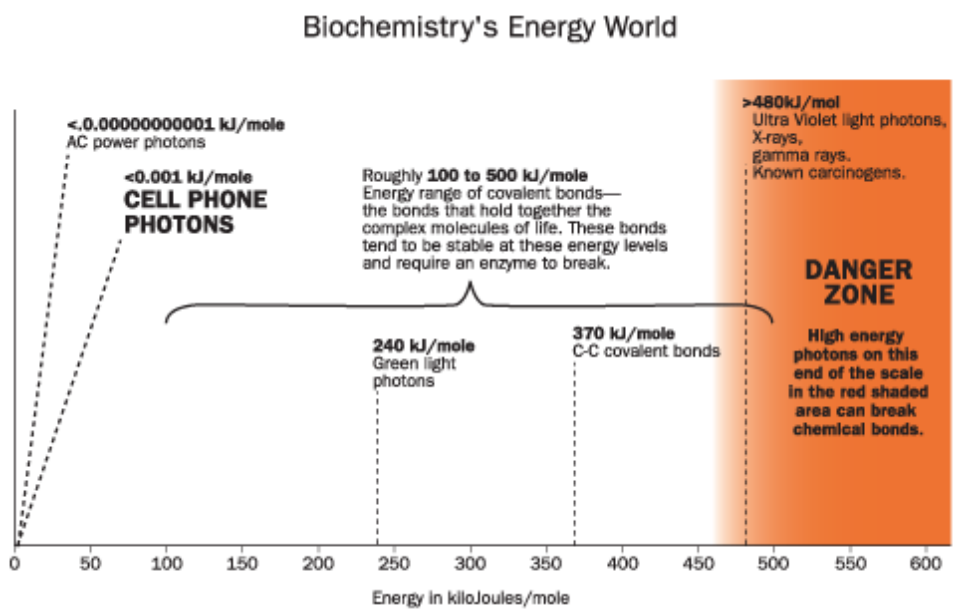

**A chart simplified from the printed version of this article in *Skeptic* magazine Vol. 15, No. 4.** This *eSkeptic* version says that the brain receives only a tiny amount of energy from a cell phone compared to that generated by normal activity such as working out. The body's powerful temperature control system deals with this extra energy without breaking into a sweat. The *Skeptic* magazine article compares the energy required to break the chemical bonds in living cells with the energy level of cell phone photons and other forms of electromagnetic energy. The result is the same. Cell phones cannot damage living tissue or cause cancer.

All physicists admit that we do not understand everything, and we may never. After more than half a century of vigorous study, I have made no progress toward an answer to that age-old question, "What do women want?" More research is necessary.

Physicists have solved the problem of microwave radiation and absorption. We know exactly what happens to the radiation, and there is no fuzzy area about it that we do not understand. The epidemiologists hear instead that physicists do not know of a mechanism by which the radiation might cause cancer.

The epidemiologists explain away their great discovery that cell phones protect against cancer and suspect that they may cause brain cancer because they believe the first has no mechanism and the second may have an unknown one. I argue strongly that there is no possible mechanism, known or unknown, by which cell phone radiation might cause cancer. However, the epidemiologists are wrong that there is no way by which cell phones might reduce the risk of brain cancer.

Here is my proposal. When our brains absorb energy from cell phones, there is a small temperature increase. When our bodies wish to energize our defense systems and to discomfit the bad guys, the immune system raises the temperature. If the problem is local, the innate immune system produces inflammation. If the problem is general, the innate immune system produces fever. Evidently, a slight, but noticeable temperature increase is beneficial to us.